\documentclass[letterpaper,12pt]{article}   
\usepackage{osajnl2} 
\usepackage[draft]{hyperref} 
\begin{document}

\title{Wigner function of pulsed fields by direct detection}
\author{Maria Bondani,$^{1,*}$ Alessia Allevi,$^2$ and Alessandra Andreoni$^3$}
\address{$^1$ National Laboratory for Ultrafast and Ultraintense
Optical Science - C.N.R.-I.N.F.M.,\\ via Valleggio, 11, I-22100, Como, Italy}
\address{$^2$ C.N.I.S.M., U.d.R. Milano,\\ via Celoria, 16, I-20133, Milano, Italy}
\address{$^3$ Dipartimento di Fisica e Matematica, Universit\`a degli Studi dell'Insubria and C.N.I.S.M., U.d.R. Como,\\ via Valleggio, 11, I-22100, Como, Italy}
\address{$^*$Corresponding author: maria.bondani@uninsubria.it}
\begin{abstract}
We present the reconstruction of the Wigner function of some classical pulsed optical states obtained by direct measurement of the detected-photon probability distributions of the state displaced by a coherent field. We use a photodetector endowed with internal gain, which is operated in the non-photon-resolving regime. The measurements are performed up to mesoscopic intensities (up to more than 30 photons per pulse). The method can be applied to characterize nonclassical continuous-variable states.
\end{abstract}

\begin{figure}[htbp]
  \centering
  \includegraphics[angle=90,width=0.44\textwidth]{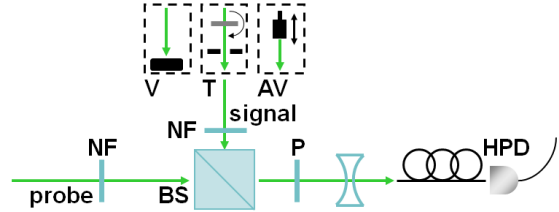}
  \caption{(Color online) Scheme of the experimental setup. HPD, hybrid photo-detector; BS, beam splitter; NF, neutral density filter; P, polarizer. Components in dotted boxes are alternately activated to produce different signal states: V, vacuum state; T, single-mode thermal state; AV, phase-averaged}\label{setup}
  \end{figure}

\begin{figure}[htbp]
  \centering
  \includegraphics[width=0.44\textwidth]{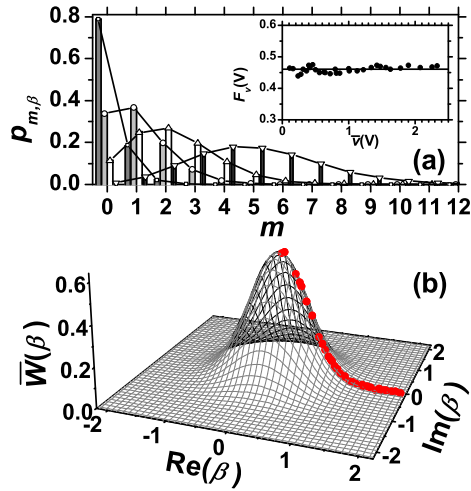}
  \caption{(Color online) Vacuum state. (a) Inset: Fano factor, $F_v$, as a function of $\bar{v}$. Main: reconstructed $p_{m,\beta}$ distributions at different $|\beta|$-values (bars) and theoretical curves (symbols+lines); (b) experimental reconstruction of a section of the Wigner function (dots) superimposed to the theoretical surface.} \label{vacuum}
  \end{figure}

\begin{figure}[htbp]
  \centering
  \includegraphics[width=0.44\textwidth]{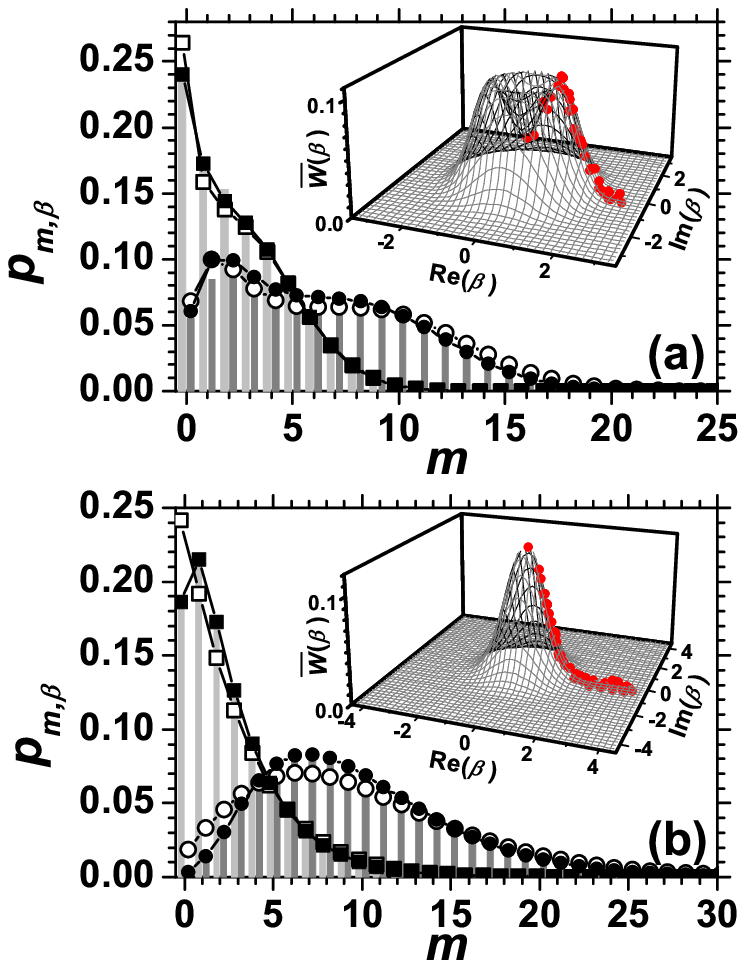}
  \caption{(Color online) (a) Phase-averaged coherent state. Main: reconstructed $p_{m,\beta}$ distributions for two values of $|\beta|$ (bars) and theoretical curves corrected for the overlap parameter (black symbols + line) or not (white symbols + line). Inset: experimental reconstruction of a section of the Wigner function (dots) superimposed to the theoretical surface; (b) same as (a) for the thermal state.} \label{phaseinsens}
  \end{figure}

The effective implementation of quantum communication protocols requires the generation and measurement of optical states in the pulsed domain. In the continuous variable regime, the characterization of pulsed states of light is usually provided by time-resolved homodyne tomography (HT) \cite{teo,sper,rassegna_OMODINA}, which allows obtaining, for instance, the Wigner function of the state \cite{Wigner}. Though HT has proved its effectiveness on classical and quantum states further optimizations to achieve a proper spatio-temporal mode matching of signal and local oscillator \cite{rassegna_OMODINA} are needed in the case of pulsed fields.
On the basis of the results derived in \cite{Cahill}, it was demonstrated that also direct detection can be used as an alternative to HT to obtain the Wigner function starting from photon-number distributions \cite{Vogel96,Banas96}. The only experimental work \cite{Banas99} was performed with continuous-wave weak fields and single-photon avalanche detectors.
The method consists in detecting the light exiting a beam-splitter that mixes the signal with a coherent probe field, whose amplitude and phase can be varied (see Fig.~\ref{setup}). The Wigner function of the state can be written as
\begin{eqnarray}
 W(\alpha) = \frac{2}{\pi} \sum_{n=0}^\infty (-1)^n p_{n,\alpha}\; ,
\label{eq:wigner}
\end{eqnarray}
where $p_{n,\alpha}$ is the photon-number distribution of the field displaced by $\alpha$. To obtain $p_{n,\alpha}$ reconstruction/inversion methods must be implemented \cite{Allevi} that require a known and sufficiently high photon-detection efficiency \cite{Kiss}.
%
%
However, in some situations, knowing the statistics of detected photons can be sufficient to characterize the signal field state, as demonstrated in the pioneering work of Arecchi \cite{Arecchi}. In fact, if we are able to characterize the detector response, we can infer the photon-number distribution from that of detected photons without inversion. Obviously, obtaining or not the statistics of detected photons depends on the photon-counting capability of the detector, a fact that limits the range of intensities that can be investigated.\\
In this Letter we exploit the linear response of a photodetector endowed with high gain in the scheme in Fig.~\ref{setup} to directly reconstruct the Wigner function of some pulsed classical states of light in the mesoscopic domain. The detector output is analyzed through a new method \cite{selfconsistent} that allows obtaining the detected photon distribution without any \emph{a priori} knowledge of the states \cite{ASL}. We model the detection process as a Bernoullian convolution \cite{Mandel64,Kelley64} and the overall amplification/conversion process through a very precise factor, which is taken as constant, $\gamma$. Our method has the advantage of being self-consistent as the value of $\gamma$ is obtained from measurements on the very field under investigation \cite{selfconsistent}. The idea is to measure a field state at different values of the overall detection efficiency of the apparatus, $\eta$, and to calculate the Fano factor, $F_v=\sigma^2(v)/\overline{v}$, of the output voltages, $v$. Since we can write $F_v  = (Q/\overline{n})\overline{v} + \gamma$ ($Q$ is the Mandel parameter), $\gamma$ can be obtained by fitting the experimental $F_v$ versus $\overline{v}$ data as all the dependence on the field is in the slope, $Q / \bar{n}$.
Once $\gamma$ is evaluated, we find the detected-photon distribution by dividing $v$ by $\gamma$ and re-binning in unitary bins.\\
Let us call $p_{m,\beta}$ the distribution of detected photons, with $\beta=\sqrt{\eta}\alpha$ the detected amplitude of the displacement field. If $p_{m,\beta}$ is linked to $p_{n,\alpha}$ by a Bernoullian convolution, we can write an expression analogous to Eq.~(\ref{eq:wigner}),
$\overline{W}(\beta) =2/\pi \sum_{m=0}^\infty (-1)^m p_{m,\beta}$, where
\begin{eqnarray}
 \overline{W}(\beta) =\frac{2}{\pi (1-\eta)}\int d^2\beta' e^{-\frac{2}{1-\eta}|\beta-\beta'|^2} W(\beta'/\sqrt{\eta})\; .
\label{eq:wigLOSS}
\end{eqnarray}
is the Wigner function in the presence of losses \cite{Banas96} and $W(\beta/\sqrt{\eta})=W(\alpha)$ is that of the photons.
It can be demonstrated that, for classical states such as non-squeezed gaussian states and their linear superpositions, we have $\overline{W}(\beta)=W(\alpha)$, \textit{i.e.} the integral in Eq.~(\ref{eq:wigLOSS}) preserves the functional form of the Wigner distribution.\\
In the scheme in Fig.~\ref{setup}, a frequency-doubled Q-switched Nd:YAG laser at 15~kHz repetition rate (Quanta System) provides linearly polarized pulses at 532~nm of $\sim200$~ns duration. The beam is spatially filtered and split into two parts serving as signal and probe fields. The photon-number statistics of the probe is left Poissonian while that of the signal is modified in order to get the different states to be measured. Signal and probe are then mixed at a cube beam-splitter and a portion of the mixed field is delivered to a hybrid photodiode module (HPD, H8236-40 with maximum quantum efficiency $\sim0.4$ at 550~nm, Hamamatsu) through a divergent lens and a multimode optical fiber (600~$\mu$m core-diameter). HPD has limited photon-number resolving capability but is linear over a wide intensity range.
%
%
Intensity and phase of the probe field are modified by a variable neutral density filter and by a piezoelectric movement, respectively.\\
We describe in detail the reconstruction of the Wigner function of the vacuum state, which was obtained by simply blocking the signal.
As the state is phase-independent, its Wigner function is rotationally invariant about the origin of the phase space and the reconstruction of a section is enough for full characterization: we thus only varied the probe amplitude $|\beta|$. We implemented the characterization procedure described above at each value of $|\beta|$ by changing $\eta$ by means of a polarizer set in front of the optical fiber (we estimate a maximum value of $\eta\sim0.31$). We recorded $N = 30000$ laser shots for each $\eta$-value. The Fano factor for this series of data is shown in the Inset of Fig.~\ref{vacuum}(a): as expected for Poissonian light, the angular coefficient is virtually zero. Examples of the obtained $p_{m,\beta}$ are shown in Fig.~\ref{vacuum}(a) as bars. The agreement with the theoretical Poissonian distributions (see Fig.~\ref{vacuum}(a), symbols+line), calculated with the measured mean values, can be checked by evaluating the fidelity $f_p =\sum_{m=0}^M p_m^{\rm th}p_m^{\rm sp}$, in which
$M$ is the number of elements of the distribution experimentally reconstructed. For the data shown in Fig.~\ref{vacuum}(a), we have $f \geq 0.9999$.
By using the measured $p_{m,\beta}$  we obtain the experimental $\overline{W}(\beta)$ displayed in Fig.~\ref{vacuum}(b) as dots. The experimental data perfectly lie on a section of the theoretical surface as calculated from the general expression for a single-mode non-squeezed Gaussian state \cite{nap04},
\begin{eqnarray}
 \overline{W}(\beta) = \frac{2}{\pi} \frac{1}{2m_{\rm th}+1} \exp(-\frac{2|\beta - \beta_0|^2}{2m_{\rm th}+1})\; ,
\label{eq:wignergauss}
\end{eqnarray}
in the case of no thermal photons ($m_{\rm th}=0$) and no coherent displacement ($\beta_0=0$).
For a quantitative estimation of the quality of the reconstruction, we evaluate the mean error $\epsilon = \sum_{k=1}^K [\overline{W}^{\rm th}(\beta_k)-\overline{W}^{\rm sp}(\beta_k)]/N$. The presented data yield $\epsilon= -8.4\times10^{-4}$.\\
To enforce the usefulness of our method, we consider two further cases in which phase-independent signal fields enter the BS: (a) a phase-averaged coherent state, obtained by randomizing the relative phase between signal and probe with a piezoelectric movement operated at a frequency of $\sim 300$~Hz and covering 1.2~$\mu$m span, and (b) single-mode thermal state, obtained by inserting a rotating ground glass plate into the pathway of the signal field, followed by a pin-hole selecting a single speckle.
In the main panels of Fig.~\ref{phaseinsens} we present the $p_{m,\beta}$ (bars) obtained from two sets of data corresponding to different $|\beta|$ values.
%
%
The reconstructed $p_{m,\beta}$ for selected values of $|\beta|$ were then used to recover a section of the Wigner function (dots in the Insets of Fig.~\ref{phaseinsens}).\\
To compare the experimental results with theory, we have to take into account the mode-mismatching between probe and signal. In fact, the imperfect superposition at BS leaves some residual probe light that is measured together with the mixed state. Thus the measured distributions result to be the convolution of those of the mixed state with the Poissonian distribution of the residual probe, while the Wigner function is the product of the expected Wigner function with that (Gaussian) of the residual probe \cite{Banas99}. In case (a) the overlap between signal and probe was $\sim 91\%$ while in case (b) it was $\sim 65\%$, owing to the further difficulty of producing a stable single-mode thermal state.\\
In Fig.~\ref{phaseinsens} we plot two theoretical curves superimposed to each set of experimental $p_{m,\beta}$. One (white symbols+line) was evaluated from the expected theoretical distribution in which the measured mean values for signal and probe were inserted. The other (black symbols+line) was evaluated by calculating the expected theoretical distribution with a probe reduced by a factor  corresponding to the overlap and then convolving it with a Poissonian distribution having a mean value equal to that of residual probe. The latter distributions exhibit better superimposition, also testified by the increase in the fidelity from $f\sim 0.998$ to $f\sim 0.999$ in (a) and from $f\sim 0.990$ to $\sim 0.999$ in (b). Finally, in the Insets of Fig.~\ref{phaseinsens} we plot the theoretical Wigner functions obtained upon the correction procedure described above. The non-corrected expected surfaces are not plotted for the sake of clarity. The expected Wigner function of the phase-averaged coherent state is
\begin{eqnarray}
 \overline{W}(\beta) = \frac{2}{\pi} I_0(4|\beta||\beta_0|)\exp[-2(|\beta|^2+|\beta_0|^2)]\; ,
\label{eq:wignerphav}
\end{eqnarray}
in which $I_0(x)$ is the modified Bessel function and $|\beta_0|^2$ is the mean number of detected photons in the signal at the output of the BS. The Wigner function of the single-mode thermal state comes from Eq.~(\ref{eq:wignergauss}) for $\beta_0=0$ and $m_{\rm th}$ equal to the number of detected thermal photons in the signal field at the output of the BS.
Since during the experiments we had some fluctuations of the signal field intensity, we kept the value of the signal as a fitting parameter and compared it with the average of the measured values. For the two cases under investigation we got $|\beta_0|^2=1.41$ and $m_{\rm th}=1.96$ to be compared with 1.55 and 1.63, respectively. The errors on the Wigner were $\epsilon=-7.8\times10^{-5}$ and $\epsilon= 4.1\times10^{-4}$.\\
In conclusion, we implemented the reconstruction of the Wigner function of some pulsed classical light fields starting from direct measurements of the statistics of the detected photons. What made the experiments feasible was the choice of a single-photon sensitive linear detector together with a new data analysis method. Works are in progress at our Laboratory in Como, whose aim is the reconstruction of a phase-dependent signal field, namely a coherent field, as well as to characterize non-classical states, both squeezed and conditional, in the same intensity regime we explored for classical ones.

The Authors thank M. G. A. Paris (Milano University) for fruitful discussions and P. Salvadeo and A. Agliati (Quanta System, Italy) for the long-term loan of the laser and for the promptness of their technical assistance.


\clearpage

\section*{List of Figure Captions}

Fig. 1. (Color online) Scheme of the experimental setup. HPD, hybrid photo-detector; BS, beam splitter; NF, neutral density filter; P, polarizer. Components in dotted boxes are alternately activated to produce different signal states: V, vacuum state; T, single-mode thermal state; AV, phase-averaged coherent state.
\clearpage
\noindent
Fig. 2. (Color online) Vacuum state. (a) Inset: Fano factor, $F_v$, as a function of $\bar{v}$. Main: reconstructed $p_{m,\beta}$ distributions at different $|\beta|$-values (bars) and theoretical curves (symbols+lines); (b) experimental reconstruction of a section of the Wigner function (dots) superimposed to the theoretical surface.
\clearpage
\noindent
Fig. 3. (Color online) (a) Phase-averaged coherent state. Main: reconstructed $p_{m,\beta}$ distributions for two values of $|\beta|$ (bars) and theoretical curves corrected for the overlap parameter (black symbols + line) or not (white symbols + line). Inset: experimental reconstruction of a section of the Wigner function (dots) superimposed to the theoretical surface; (b) same as (a) for the thermal state.

\begin{thebibliography}{99}
\bibitem{teo} U. Leonhardt, H. Paul, and G. M. D'Ariano, ``Tomographic reconstruction of the density matrix via pattern functions'', Phys. Rev. A \textbf{52}, 4899 (1995).
\bibitem{sper} D. T. Smithey, M. Beck, M. G. Raymer, and A. Faridani, ``Measurement of the Wigner distribution and the density matrix of a light mode using optical homodyne tomography: Application to squeezed states and the vacuum'', Phys. Rev. Lett \textbf{70}, 1244 (1993).
\bibitem{rassegna_OMODINA} A. I. Lvovsky and M. G. Raymer, ``Continuous-variable optical quantum state tomography'', to appear in Rev. Mod. Phys,  quant-ph/0511044v2.
\bibitem{Wigner} M. Hillery, R. F. O'Connell, M. O. Scully and E. P. Wigner, ``Distribution functions in physics: Fundamentals'', Phys. Rep. \textbf{106}, 121 (1984).
\bibitem{Cahill} K. E. Cahill and R. J. Glauber, ``Density Operators and Quasiprobability Distributions'', Phys. Rev. \textbf{177}, 1882 (1969).
\bibitem{Vogel96} S. Wallentowitz and W. Vogel, ``Unbalanced homodyning for quantum state measurements'', Phys. Rev. A \textbf{53}, 4528 (1996).
\bibitem{Banas96} K. Banaszek and K. W$\acute{\mathrm{o}}$dkiewicz, ``Direct Probing of Quantum Phase Space by Photon Counting'', Phys. Rev. Lett. \textbf{76}, 4344 (1996).
\bibitem{Banas99} K. Banaszek, C. Radzewicz, K. W$\acute{\mathrm{o}}$dkiewicz, and J. S. Krasi$\acute{\mathrm{n}}$ski, ``Direct measurement of the Wigner function by photon counting'', Phys. Rev. A \textbf{60}, 674 (1999).
\bibitem{Allevi} A. Allevi, A. Andreoni, M. Bondani, G. Brida, M. Genovese, M. Gramegna, S. Olivares, M. G. A. Paris, P. Traina, and G. Zambra, ``State reconstruction by simple measurements'', quant-ph/0903.0104.
\bibitem{Kiss} T. Kiss, U. Herzog, and U. Leonhardt, ``Compensation of losses in photodetection and in quantum-state measurements'', Phys. Rev. A \textbf{52}, 2433 (1995).
\bibitem{Arecchi} F. T. Arecchi, ``Measurement of the Statistical Distribution of Gaussian and Laser Sources'', Phys. Rev. Lett. \textbf{15}, 912 (1965).
\bibitem{selfconsistent} M. Bondani, A. Allevi, A. Agliati, and A. Andreoni, ``Self-consistent characterization of light statistics'', J. Mod. Opt. \textbf{56}, 226 (2009).
\bibitem{ASL} M. Bondani, A. Allevi, and A. Andreoni, ``Light statistics by non-calibrated linear photodetectors'', Adv. Sci. Lett. (in press) and quant-ph/0810.4026.
\bibitem{Mandel64} L. Mandel, E. C. G. Sudarshan, and E. Wolf, ``Theory of photoelectric detection of light fluctuations'', Proc. Phys. Soc. (London) \textbf{84}, 435 (1964).
\bibitem{Kelley64} P. L. Kelley, and W. H. Kleiner, ``Theory of electromagnetic field measurement and photoelectron counting'', Phys. Rev. \textbf{136}, A316 (1964).
\bibitem{nap04} G. S. Agarwal, ``Wigner-function description of quantum noise in interferometers'', J. Mod. Opt. \textbf{34}, 909 (1987).
\end{thebibliography}
\end{document}